\RequirePackage{ifpdf}
\ifpdf 
\documentclass[pdftex]{sigma}
\else
\documentclass{sigma}
\fi

\usepackage{mathrsfs}

\numberwithin{equation}{section}

\begin{document}

\allowdisplaybreaks

\renewcommand{\PaperNumber}{023}

\FirstPageHeading

\ShortArticleName{Nonlinear Dirac Equations}

\ArticleName{Nonlinear Dirac Equations}

\Author{Wei Khim NG and Rajesh R. PARWANI}
\AuthorNameForHeading{W.K. Ng and R.R. Parwani}
\Address{Department of Physics, National University of Singapore, Kent Ridge, Singapore}
\Email{\href{mailto:g0500437@nus.edu.sg}{g0500437@nus.edu.sg}, \href{mailto:parwani@nus.edu.sg}{parwani@nus.edu.sg}}
\URLaddress{\url{http://staff.science.nus.edu.sg/~parwani/}}

\ArticleDates{Received November 12, 2008, in f\/inal form February 23,
2009; Published online February 27, 2009}

\Abstract{We construct nonlinear extensions of Dirac's relativistic electron equation that preserve its other desirable properties such as
locality, separability, conservation of probability and Poincar\'{e} invariance.
We determine the constraints that the nonlinear term must obey and
classify the resultant non-polynomial nonlinearities in a double
expansion in the degree of nonlinearity and number of derivatives.
We give explicit examples of such nonlinear equations, studying
their discrete symmetries and other properties. Motivated by
some previously suggested applications we then consider nonlinear terms that simultaneously
violate Lorentz covariance and again study various explicit
examples. We contrast
our equations and construction procedure with others in the
literature and also show that our equations are not gauge equivalent to the linear Dirac equation. Finally we outline various physical applications for these equations.}


\Classification{81P05; 81Q99; 83A05}

\section{Introduction}\label{section1}

When Schr\"{o}dinger obtained his wave equation to realise de
Broglie's speculation about the wave nature of particles, he used
a number of heuristic arguments and assumed the simplest
possibility, that of linearity of the equation~\cite{Sch}. Fortunately that
assumption led to very good agreement with experiment and till
today no deviations from quantum linearity have been detected eventhough a few low energy experiments have attempted to observe
them \cite{Expts,Expts1,Expts2,Expts3}. Currently the main interest in nonlinear Schr\"{o}dinger
equations is that they appear, in form, as approximations in
optics and condensed matter \cite{book1,book2}.

When Dirac generalised Schr\"{o}dinger's equation to the
relativistic domain, he too kept lineari\-ty. Nonlinear versions of
Dirac's equation have been studied for various purposes since
then. Heisenberg's proposal \cite{Hei} was in the context of f\/ield theory and
was motivated by the question of mass. In the quantum mechanical
context, nonlinear Dirac equations have been used as ef\/fective
theories in atomic, nuclear and gravitational
physics \cite{Misc,Misc1,Misc2}. Some of the simpler versions have been analysed
rigorously \cite{Este}.

Although there is as yet no evidence for fundamental quantum
nonlinearities, their absence is seen as a puzzle by  several
authors and requires an understanding \cite{Bial,kibble,wein1,doe,RP1}. Based on an extrapolation
of some information theoretic arguments at the non-relativistic
level, it was proposed in \cite{RP2} that perhaps quantum
linearity might be intimately tied to Lorentz invariance and that
the possible violation of the latter at a fundamental level might
lead to quantum nonlinearities. If true, then perhaps the
appropriate regime to seek such inter-related violations would be
at high energies or at very short distances.

Since quantum nonlinearities, if they exist, must be very small,
the best place to search for them is where they might show up at
leading order, not masked by other corrections. Thus one hopes to
detect the nonlinearities at the quantum mechanical level, rather
than as supplements to loop ef\/fects in f\/ield theory. Neutrinos are
therefore an ideal probe of such potential nonlinearities as they
are weakly interacting and so not af\/fected much by f\/ield theory
corrections.

Indeed, neutrino oscillations were suggested in \cite{RP2} as one
place where quantum nonlinearities might be relevant and a
heuristic study was conducted using a provisional nonlinear Dirac
equation. That equation was very complicated and it did not
conserve probability.

In this paper we discuss  Dirac equations, {\it at the quantum mechanical level},
which preserve all the others desirable features such as
conservation of probability. We intend to use these equations to
study not just neutrino oscillations but also various other
high-energy phenomenon which are brief\/ly discussed in the last
section.

However it is possible that our equations might also be relevant
as approximate equations, for use either in particle physics
or condensed matter physics, and we discuss this also in the
concluding section.

As there are various obstacles to generalising the
non-relativistic information theory approach of \cite{RP2} to the relativistic domain, we
proceed in a dif\/ferent manner here. We write the nonlinear
equation as
\begin{gather}
        \left(i\gamma^\mu\partial_\mu-m+F\right)\psi=0 , \label{nld}
    \end{gather}
where $F$ is a function of the wavefunction $\psi$, its adjoint and their
derivatives\footnote{But we do not consider $F$'s that have free
derivatives acting to the right on the f\/inal $\psi$ of the
equation (\ref{nld}). So our nonlinearity is a matrix in spinor space with spacetime dependent coef\/f\/icients.}. We begin by requiring, just as for $F=0$,
that equation (\ref{nld}) be local, Poincar\'{e} covariant,
conserves probability and is separable for multi-particle states. The constraints on $F$ are
then solved in an expansion procedure to be detailed in Section~\ref{section2.5.1}. That is, we implement a systematic scheme to construct
a large class of nonlinear extensions of the Dirac equation.

The constraints we adopt are similar to those used in
understanding non-relativistic quantum theory in \cite{RP3,RP3a}.
There it was deduced that the Schr\"{o}dinger equation is the
unique single universal parameter ($\hbar$) extension of classical
ensemble dynamics. Although the speed of light,~$c$, is a
universal parameter for relativistic dynamics, it  already appears
at the classical level and plays the role of converting the
dimensions of space to those of time. One expects that further
extensions of quantum theory either at the non-relativistic or
relativistic level would involve other universal parameters, for
example a universal length.

Our approach and most of our results dif\/fer from previous
constructs of nonlinear Dirac equations in the literature. Most
studies \cite{Misc,Misc1,Misc2} do not impose separability, which is a~strong constraint
that leads to non-polynomiality of~$F$. In \cite{doebner}
separability was imposed in a~somewhat dif\/ferent manner from what
we do here, but more importantly the authors of \cite{doebner}
only considered nonlinear Dirac equations that are obtained from
the linear Dirac equation through a process of gauge-completion:
thus their class of equations is more restrictive than ours. Some
further contrasts of our procedure compared to others \cite{NLDS,NLDS1} is
that {\it we allow derivatives of the wavefunction in~$F$}, and also study
nonlinearities which violate one or more of the discrete~$\mathcal{P}$,~$\mathcal{C}$,~$\mathcal{T}$ symmetries as such cases are
expected to be phenomenologically relevant.

Furthermore, proceeding with the suggestion of \cite{RP2}, we also
construct versions of (\ref{nld}) that are simultaneously  Lorentz
violating and nonlinear: such equations have also not been studied before in general; however we note that one example of such an equation, without derivatives in the nonlinearity, has been studied in~\cite{bogo1,bogo1a}, motivated by anisotropic space-times \cite{bogo2}.

The rest of the paper is structured as follows: In Section~\ref{section2} we
discuss and make explicit the various constraints on the nonlinear
term $F$; we note that the class of nonlinearities we consider can
also be motivated without imposing separability and so are
potentially useful also as ef\/fective equations at low energies.
The simplest examples of such equations are discussed in Section~\ref{section3}
followed by their plane-wave solutions and the corresponding
dispersion relations in Section~\ref{section4}. In Section~\ref{section5} we study examples
of~$F$ that simultaneously violate Lorentz covariance. In Section~\ref{section6} we illustrate more complicated examples of the nonlinear
equations and also discuss the alternative approach whereby the
nonlinear equations are obtained from a Lagrangian. In Section~\ref{section7}
we explain how to distinguish our nonlinearities from those that
may be obtained from the linear equation through a nonlinear gauge
transformation. A summary and outlook is in Section~\ref{section8}.

Although the evolution equation (\ref{nld}) has been modif\/ied, we keep the usual kinematical structure of quantum mechanics; some arguments, that fundamental nonlinear
quantum theories are intrinsically pathological, are discussed in the f\/inal section.
The conventions we use are similar to those in the textbook~\cite{IZ}; unless stated, our discussion is representation independent.
Although we work in $3+1$ dimensional f\/lat spacetime with metric $g^{\mu\nu}=(1,-1,-1,-1)$, some ef\/fects of gravity could possibly be encoded in an ef\/fective nonlinearity; we do not study in this paper explicit couplings to gravity though this might yield some interesting consequences as seen for the linear Dirac equation~\cite{kost2}.

\section{Constraints}\label{section2}

The usual, linear, quantum-mechanical (``f\/irst quantised") Dirac equation has many appealing properties
which we will mostly preserve so as to achieve a minimal
deformation. Later in Sections~\ref{section5} and~\ref{section6}, we discuss the possibility
of further extensions motivated by physical considera\-tions.

We now list and explain the various constraints that we are going
to impose on the nonlinear Dirac equation and hence on $F$ in~(\ref{nld}).

    \subsection{Locality}\label{section2.1}

     We continue to assume that physics, as described by the wavefunction $\psi$, is accurately captured by a local evolution equation: that is we require $F$ to depend only on $\psi$, its conjugate and their derivatives all evaluated at a single point~$x$. Note that $F$ below is in general a {\it matrix} in spinor space though later we will specialise to various cases, such as $F$ proportional to the identity matrix.

     Notice that we demand locality of the equations of motion rather than of a Lagrangian. This means that some of our equations might not be obtainable from a local Lagrangian. One could of course implement a construction procedure similar to that described below at the local Lagrangian level: we illustrate this in Section~\ref{section6} and discuss the relative advantages and disadvantages.

    \subsection{Poincar\'{e} invariance}\label{section2.2}

    Under the Poincar\'{e} transformation $x'=\Lambda x+a$
    the linear Dirac equation is covariant if the wavefunction transforms as \cite{IZ}
    \begin{gather*}
            \psi'(x') = S(\Lambda)\psi(x)=\psi\left(\Lambda^{-1}(x'-a)\right),
    \end{gather*}
    where $S^{-1}(\Lambda)\gamma^\mu{\Lambda^\nu}_\mu
        S(\Lambda)=\gamma^\nu$. Explicitly we have
$S(\Lambda) =\exp({-\frac{i}{4}\sigma_{\alpha\beta}\omega^{\alpha\beta}})$,
with $\omega^{\alpha\beta}$ the transformation parameters.
If we demand that the nonlinear equation~(\ref{nld}) be covariant under the same transformations then we obtain the following constraint,
 \begin{gather*}
            S^{-1}(\Lambda)F'S(\Lambda)=F   , 
\end{gather*}
        where $F'$ is the Poincar\'{e} transformed $F$; recall that
        $F$ is a function depending on $\bar \psi$, $\psi$ and their derivatives.

    \subsection{Hermiticity}\label{section2.3}

    In quantum mechanics we usually require the Hamiltonian to be Hermitian so as to guarantee reality of eigenvalues. Rewriting the nonlinear Dirac equation in Hamiltonian form we have,
        \begin{gather*}
               i\frac{\partial}{\partial t}\psi = \left(H_D-\beta F\right)\psi,
    \end{gather*}
    where $\beta=\gamma^0$ and $H_D$ is 
    the linear Dirac Hamiltonian. Since $H^\dag_D=H_D$, thus we also impose\footnote{Recall, we are adopting the standard kinematical structure of quantum mechanics, in particular the standard inner product. See also the f\/irst footnote.}
        \begin{gather}
                \gamma^0 F^\dag\gamma^0 = F   . \label{herm}
    \end{gather}

      \subsubsection{Current conservation}\label{section2.3.1}

    In terms of the familiar adjoint $\bar{\psi} = \psi^{\dag} \gamma^{0}$, the linear Dirac equation has the conserved current
\begin{gather}
j^\mu=\bar \psi\gamma^\mu\psi, \label{curr}
\end{gather}
which allows $\psi^{\dag} \psi$ to be interpreted as a probability density. The divergence of the same expression (\ref{curr}) in the nonlinear theory is
       \begin{gather}
            \partial_\mu j^\mu  = \bar \psi\big(iF-i\gamma^0F^\dag\gamma^0\big)\psi, \label{leak}
        \end{gather}
  which vanishes due to the Hermiticity  condition (\ref{herm}).

    Thus requiring Hermiticity of the Hamiltonian also ensures conservation of (\ref{curr}). On the other hand, in some future applications, we may want to consider non-Hermitian Hamiltonians that model open systems. Then the right-hand-side of (\ref{leak}) can be used to measure leakage from the system.

\subsubsection{Chiral current}\label{section2.3.2}

For completeness we also discuss the chiral current, for which the  expression in the linear theory is
$j^\mu_5 = \bar \psi\gamma^\mu\gamma_5\psi$. Using the nonlinear equations of motion, we obtain
        \begin{gather*}
            \partial_\mu j^\mu_5 = -i\bar
            \psi\big(\gamma^0F^\dag\gamma^0\gamma_5+\gamma_5F\big)\psi + 2im \bar \psi \gamma_5 \psi.
        \end{gather*}
        For the usual chiral current to be conserved in the massless, $ m \to 0$, limit of the nonlinear equation, we require
\begin{gather*}
            \gamma_5F+\gamma^0F^\dag\gamma^0\gamma_5=0  ,
\end{gather*}
which, on using the Hermiticity condition (\ref{herm}), simplif\/ies
to
\begin{gather*}
\{F, \gamma_5\} =0   .
\end{gather*}

  \subsection{Universality}\label{section2.4}

The usual Dirac equation has the property, as all linear equations
do, that it is invariant under a rescaling of the wavefunction,
$\psi \to \lambda \psi$. In quantum mechanics such a condition
allows solutions of the equation to be freely normalised, which
is not only convenient but also sometimes demanded for an
interpretation of measurements \cite{Bial,kibble,wein1,doe}.

We would like our nonlinear generalisation to preserve the same
scale-invariance property, which one may motivate with alternative
reasoning as follows. We desire equations that are as universal as
possible. So, for example, the equation should have the same form
whether it describes a single particle or a system of particles.
More specif\/ically, the parameters describing the strength of the
nonlinearity $F$ should not be dependent on the number of
particles in the system, just as Planck's constant $\hbar$ is
universal in the multiparticle Schr\"{o}dinger equation.

If $\psi$ represents the wavefunction for a $N$-particle state, then the normalisation of probability implies that
the dimension of $\psi$ depends on $N$, just as in the non-relativistic case \cite{RP3,RP3a}, and so the dimension of $F$ would then be $N$ dependent in general. We can avoid this conclusion by requiring that $F$ have the above-mentioned scaling property
\begin{gather}
F(\lambda \psi) = F (\psi), \label{scale}
\end{gather}
where we mean that the wavefunction and its conjugate are all scaled by the same factor $\lambda$ on the left-hand-side.
Equation (\ref{scale}) implies that $F$ must be non-polynomial,
\begin{gather}
F \sim F(A/B),  \label{nonpoly}
\end{gather}
where $A$, $B$ have equal factors of the wavefunction.

    \subsection{Separability}\label{section2.5}

    The usual Dirac equation may be used to describe a collection of particles and is separable for independent subsystems. It seems useful to have this separability property also for our nonlinear generalisation. However as we will explain in a later section, one may omit the separability constraint in favour of other arguments which result in similar forms for the eventual $F$'s, and those forms anyway become separable with a suitable interpretation of the multiparticle states. Thus with the same structure for $F$ we can use the equation for fundamental, phenomenological or ef\/fective dynamics.

Let us review separability f\/irst for the linear Dirac equation so as to motivate suitable def\/i\-nitions of $\bar
    \psi$ and $j^\mu$ for many-body systems. In the multi-time formalism \cite{multi,multi1,multi2}, which preserves manifest Poincar\'{e} invariance, the many-body linear Dirac equation for non-interacting particles may be written as
    \begin{gather*}
        \sum_s
        \left(i\gamma^\mu_s\partial_{\mu,s}-m_s\right)\psi=0,
    \end{gather*}
    where
    \begin{gather*}
        \psi = \psi_1\otimes\psi_2\otimes\cdots\otimes\underbrace{\psi_s}_{s^{\rm th} \, {\rm site}} \otimes\cdots,\\
        \gamma^\mu_s = 1\otimes1\otimes\cdots\otimes\underbrace{\gamma^\mu}_{s^{\rm th} \, {\rm site}} \otimes1\otimes\cdots,  \\ 
        m_s = 1\otimes1\otimes\cdots\otimes\underbrace{m_{(s)}}_{s^{\rm th} \, {\rm site}}\otimes1\otimes\cdots
    \end{gather*}
    and  $s$ labels the particle. Consider explicitly the two-particles case,
    \begin{gather}
  \left[\left(i\gamma^\mu\partial_{\mu,1}-m_{(1)}\right)\psi_1\right]
  \otimes\psi_2+\psi_1\otimes\left[\left(i\gamma^\mu\partial_{\mu,2}-m_{(2)}\right)\psi_2\right] =  0    . \label{2.2}
    \end{gather}
Let $\phi_1$ and $\phi_2$ be arbitrary single particle wavefunctions for the two independent variables~$1$,~$2$. Then multiplying by $\bar \phi_1\otimes\bar
    \phi_2 / (\bar \phi_1 \psi_1) (\bar \phi_2 \psi_2)$ on the left
    of (\ref{2.2}), we have
    \begin{gather}\label{2.3}
        \frac{\bar{\phi_1} \left(i\gamma^\mu\partial_{\mu,1}-m_{(1)}\right)\psi_1}{\bar \phi_1 \psi_1}\otimes
        1+1\otimes\frac{\bar{\phi_2} \left(i\gamma^\mu\partial_{\mu,2}-m_{(2)}\right)\psi_2}{\bar{\phi_2} \psi_2}=0   .
    \end{gather}
The result is clearly separable in that solutions of the individual single particle Dirac equations satisfy the two-particle equation and vice-versa.

Furthermore it is easy to show that if
$\bar \psi$ for a many-body system is def\/ined to be $\bar \psi=\bar \psi_1\otimes\bar \psi_2\otimes\cdots\otimes\bar \psi_s\otimes\cdots$ then the two-particle adjoint equation that follows from (\ref{2.2}) will have the same form as the one-particle equation, and since
form-invariance is in the spirit of the universality criteria of Section~\ref{section2.4}, this justif\/ies our def\/inition.

Now consider, as an example, the expression for the multi-particle current $j^\mu$.
    Multiply (\ref{2.2}) from the left by $\bar \psi_1\otimes\bar
    \psi_2$, multiply the adjoint of (\ref{2.2}) from the right by
    $\psi_1\otimes\psi_2$, and take the dif\/ference to get
       \begin{gather}
        \big[\partial_{\mu,1}j^\mu_{(1)}\big]\otimes\bar
        \psi_2\psi_2+\bar
        \psi_1\psi_1\otimes\big[\partial_{\mu,2}j^\mu_{(2)}\big]=0\label{2.5}
    \\
       \qquad{} \Rightarrow\ \sum_{s=1}^2 \partial_{\mu,s}j^\mu_s=0,\nonumber
    \end{gather}
    where the current is def\/ined to be
    \begin{gather*}
    j^\mu_s=\bar \psi_1\psi_1\otimes\bar \psi_2\psi_2\otimes\cdots\otimes \underbrace{j^\mu_{(s)}}_{s^{\rm th}\, {\rm site}}\otimes\bar
    \psi_{s+1}\psi_{s+1}\cdots.
    \end{gather*}
    Multiplying (\ref{2.5}) by $\left(\bar \psi_1\psi_1\otimes\bar
    \psi_2\psi_2\right)^{-1}$ gives
    \begin{gather*}
        \frac{\partial_{\mu,1}j^\mu_{(1)}}{\bar \psi_1\psi_1}\otimes
        1+1\otimes\frac{\partial_{\mu,2}j^\mu_{(2)}}{\bar \psi_2\psi_2}=0   .
    \end{gather*}
    Thus conservation of individual currents implies the conservation of the two-particle current and {\it vice-versa}.

Similarly the def\/inition
\begin{gather}
    \gamma^5_s=1\otimes1\otimes\cdots\otimes\underbrace{\gamma^5}_{s^{\rm th}\,
        {\rm site}}\otimes1\otimes\cdots
\end{gather}
allows the multiparticle chiral current to be def\/ined and in the massless limit  conservation of individual chiral currents implies
the conservation of the two-particle chiral current and {\it
vice-versa}.

\subsubsection[Structure of $F$]{Structure of $\boldsymbol{F}$}\label{section2.5.1}

    We would like our nonlinear equation to be separable in this minimal sense: for a wavefunction which is the  product of two independent states, the composite equation should  decompose into two independent equations\footnote{We note that other implementations of separability might lead to more constraints, see for example~\cite{svet}.}. Looking at the expressions (\ref{2.2}), (\ref{2.3}) we see that for the nonlinear equation (\ref{nld}) to be separable as such, we require $F$ to decompose as
\begin{gather*}
F(\psi_1 \otimes \psi_2) = F(\psi_1)\otimes 1 + 1 \otimes F(\psi_2)   
\end{gather*}
for a state made up of two independent particles (constraints of this type have been studied before for non-relativistic systems in~\cite{svet1}).
Equation (\ref{nonpoly}) and the examples above suggest that this can be achieved if we have the structure{\samepage
\begin{gather*}
 F \left({\mathcal{N} \over \mathcal{D}} \right)  \sim {\mathcal{N} \over \mathcal{D}} \to { \mathcal{N}_1 \over \mathcal{D}_1}  \otimes 1 + 1 \otimes {\mathcal{N}_2 \over \mathcal{D}_2 }   .
 \end{gather*}
Thus for a product state we require $\mathcal{N} \to \mathcal{N}_1 \otimes \mathcal{D}_2 + \mathcal{D}_1 \otimes \mathcal{N}_2$ while $\mathcal{D} \to \mathcal{D}_1 \otimes \mathcal{D}_2$.}

Requiring $\mathcal{N}$ and $\mathcal{D}$ to be separately
Poincar\'{e} invariant, we see that the only functional of $\psi$
that would decompose as required for $\mathcal{D}$ is $\bar \psi
\psi$ and powers thereof. Thus our nonlinear term $F$ can be a
{\it} sum of terms of the form
    \begin{gather}
    \frac{ \mathcal{N}\left(\bar \psi,\psi\right)}{\left(\bar \psi\psi\right)^n}, \label{form1}
    \end{gather}
    subject to the other constraints that have yet to be imposed.

Our deduction of (\ref{form1}) has been somewhat heuristic and so
the reader may prefer to think of it as an {\it ansatz} within
which we discuss our equations.

As mentioned earlier, the separability condition is appropriate
for fundamental equations that describe an arbitrary collection of
particles. However if the nonlinearities are an approximate
description of an underlying dynamics, as ef\/fective equations
attempt to do, then the universality and separability arguments do
not seem appropriate. However even then one may motivate the
structure (\ref{form1}) as follows. Generally, for slowly varying
f\/ields, one may perform a gradient expansion for $F$ when seeking
local equations,
\begin{gather}
F \sim {N_0 \over D_0} + {N_1 \over D_1} + {N_2 \over D_2} + \dots
+ {N_i\over D_i}+\cdots, \label{alter}
\end{gather}
where the $N_i$'s depend on the wavefunction and contain exactly
$i$ derivatives. The $D_i$'s also depend on the wavefunction but
do not contain any derivatives.

Now in most nonlinear Scr\"{o}dinger or Dirac equations the
nonlinear terms break the scale invariance, $\psi \to \lambda
\psi$, present in the linear theory. That is, typically the
nonlinearities make the equations sensitive to the amplitude of
the f\/ields thus giving rise to very interesting phenomena. However it
is possible to have nonlinearities that preserve the scale
invariance of the linear theory and though the ef\/fects are then
likely to be milder, they can still lead be novel and interesting
ef\/fects \cite{RP4,RP4a}. So if we focus on such ``soft''
nonlinearities, and also impose Lorentz invariance, then
(\ref{alter}) is included in the form (\ref{form1}). Indeed, as we
shall verify later, even without imposing separability at the
outset, separability of the resultant structures appears to be
possible with consistent def\/initions of the multi-particle states.

In summary, we will discuss in this paper the class of
nonlinearities of the form (\ref{form1}) by looking at several
cases corresponding to a specif\/ic degree of nonlinearity,
$n=1,2,\dots$, and a~derivative expansion of the numerator.

We remark that the scale-invariant nonlinearities (\ref{form1}) we
introduce here might also be interesting for future quantum {\it field
theory} investigations: these nonlinearities correspond to
Lagrangians that are still naively power-counting renormalisable.

\subsection{Discrete symmetries}\label{section2.6}
The Standard Model of particle physics  encodes both parity and
$\mathcal{CP}$ violation as these are empirically observed facts.
Thus in our nonlinear equation we f\/ind it interesting to allow
violation of individual symmetries. However in line with general
theorems \cite{IZ,Green} on local, Hermitian, Lorentz covariant theories,  we do
f\/ind by explicit verif\/ication that our specif\/ic examples
preserve the combined $\mathcal{PCT}$ invariance although
we do not impose it.

The discrete symmetry operators are the same as in the linear theory \cite{IZ}, and they place constraints on the nonlinear term $F$ so that the nonlinear equation (\ref{nld}) is form invariant (similar to the discussion in Section~\ref{section2.2}).

Ignoring unobservable phases, the representation independent parity operator is $\hat{\mathcal{P}} = \gamma^0$ and parity invariance requires
\begin{gather*}
            \hat {\mathcal{P}}^{-1}F_P\hat{ \mathcal{P}}\equiv F   ,
        \end{gather*}
where $F_P$ is the parity transformed $F$. Charge conjugation invariance is achieved if
 \begin{gather*}
            \hat {\mathcal{C}}^{-1}F_C\hat{ \mathcal{C}}\equiv F^*,
        \end{gather*}
where $F_C$ is the charge conjugated $F$ and $\hat {\mathcal{C}} = i\gamma^2 $ in the Dirac--Pauli representation. The time-reversal invariance constraint on $F$ is
        \begin{gather*}
            \hat{ \mathcal{T}}^{-1}F_T\hat {\mathcal{T}}\equiv F^*   ,
        \end{gather*}
        $F_T$ being the time reversed $F$ and $\hat {\mathcal{T}} =i\gamma^1\gamma^3$
in the Dirac--Pauli representation\footnote{We remind the reader that we are treating the wavefunction as a classical object rather than as a Grassmann variable. Thus the denominator of (\ref{form1}) obtains a
       negative sign when performing a transpose operation, such as occurs in charge conjugation. Such negative signs mutually cancel for our scale-invariant nonlinearities~(\ref{form1}).}.

Under the combined $\mathcal{PCT}$ transformation, $\hat \Theta$,
the nonlinear Dirac equation in invariant if
\begin{gather*}
            \hat {\Theta}^{-1} F_{\Theta} \hat {{\Theta}}\equiv F,
        \end{gather*}
        where $F_{\Theta}$ is the $\mathcal{PCT}$ transformed $F$.
The representation independent form for $\hat \Theta$ is proportional to $\gamma^5$.

\section[Explicit examples of nonlinear equations with $F \propto I$, $n=1$]{Explicit examples of nonlinear equations with $\boldsymbol{F \propto I}$, $\boldsymbol{n=1}$}\label{section3}

We found earlier in Section~\ref{section2.5.1} that $F$ has the form
 \begin{gather}
    \frac{ \mathcal{N}\left(\bar \psi,\psi\right)}{\left(\bar \psi\psi\right)^n}, \label{form}
    \end{gather}
where the number of factors of the wavefunction in the numerator is $2n$.

In the absence of other dynamical f\/ields, Poincar\'{e} invariance
requires spacetime indices of matrices like $\gamma^{\mu}$ to be
contracted among themselves or with derivatives $\partial^{\mu}$.
We will assume that the spinor indices of $\psi$ and $\bar \psi$
are contracted in the natural way with $\bar \psi$ acting like a
row vector and $\psi$ a column vector, for example $ \mathcal{N}
\sim A \bar\psi B \psi C$ where $A$, $B$, $C$ are matrices in spinor
space.

In this Section we restrict the explicit discussion to the important case where $F$ is proportional to the identity matrix $I$ in spinor space,
\begin{gather}
 F = f I \label{effmass}
\end{gather}
and so the nonlinearity $f$ may be thought of as a spacetime
dependent mass. This choice is motivated by our interest in
neutrino oscillations. We also consider here only the lowest order
of nonlinearity, $n=1$. In Section~\ref{section6}, we discuss some other types
of $F$.

Current conservation for the case (\ref{effmass}) simply amounts to the statement that $f$ is a real function of the wavefunction,
\begin{gather*}
 f= f^* . 
\end{gather*}

\subsection{No derivatives}\label{section3.1}
In the absence of derivatives, the most general structure of the nonlinear term with $F \propto I$ and $n=1$ is given by
\begin{gather}
F=\frac{\bar \psi A\psi}{\bar \psi\psi}   , \label{noderone}
\end{gather}
where $A$ is a matrix. In the absence of other f\/ields which carry spacetime indices we must therefore have
\begin{gather*}
A=aI+ i b \gamma_5 , 
\end{gather*}
where $a$, $b$ are constants. The $a$ term is clearly equivalent to a
mass term in the {\it linear} equation and so may be ignored in
the following discussion. Notice that the form $A = ib \gamma_5$
in (\ref{noderone}), which is a consequence of Lorentz invariance,
also automatically satisf\/ies the  $\mathcal{PCT}$ invariance
condition.

As for individual discrete symmetries, using the equations of
Section~\ref{section2.6}, we see that the term with $b \neq 0$ preserves
$\mathcal{C}$ invariance but breaks parity. Time-reversal
invariance requires $b$ to be purely imaginary, which conf\/licts
with the requirement from current conservation which requires $b$
to be real.

We thus conclude that our simplest nonlinear equation, with $F \propto I$ and $n=1$,
\begin{gather*}
F= i \epsilon \frac{\bar \psi \gamma_5 \psi}{\bar \psi\psi}   . 
\end{gather*}
unavoidably breaks $\mathcal{P}$ and $\mathcal{CP}$, something
that is surely intriguing from the perspective of particle physics
phenomenology. We have indicated the small nonlinearity parameter
by $\epsilon$.

Note that the multiparticle version of the above equation is separable, so that does not impose additional constraints. Nonlinear Dirac equations {\it without derivatives} in the nonlinear part have been studied in~\cite{NLDS,NLDS1} and (\ref{noderone}) is a special case of the equations studied there.

\subsubsection[Lorentz vs $\mathcal{PCT}$ invariance]{Lorentz vs $\boldsymbol{\mathcal{PCT}}$ invariance}\label{section3.1.1}

Let us discuss the situation whereby the $\mathcal{PCT}$
invariance is imposed on (\ref{noderone}) f\/irst. Then we f\/ind,
using $ \hat{\Theta} \propto \gamma_5$, that we require
\begin{gather*}
[A, \gamma^5 ] =0,
\end{gather*}
which is satisf\/ied if $A$ has the form
            \begin{gather*}
                A=aI+b\gamma_5+c^{\mu\nu}\sigma_{\mu\nu}   .
            \end{gather*}
If there are no other dynamical f\/ields other than the wavefunction, then $c^{\mu \nu}$ can only be a~constant background f\/ield, thus explicitly breaking Lorentz invariance. Indeed, explicitly implementing Lorentz invariance of
(\ref{noderone}) gives
\begin{gather*}
                S(\Lambda)^{-1}AS(\Lambda)\equiv A,
            \end{gather*}
 which for the inf\/initesimal case  gives
 $\left[A,\sigma_{\alpha\beta}\right]=0$. This only allows
            \begin{gather*}
                A=aI+b\gamma_5
            \end{gather*}
as we argued earlier.

In other words, we can have $\mathcal{PCT}$ invariance even if
we give up Lorentz invariance, which again is consistent with
general results in the literature \cite{IZ,Green}.

\subsection{One derivative}\label{section3.2}

        The most general form of $F$ is now given by the linear
        combination of the following two terms,
        \begin{gather*}
            \frac{\left(\partial_\mu\bar \psi\right)A\gamma^\mu B\psi}{\bar
            \psi\psi}, \qquad \frac{\bar \psi C\gamma^\mu D\partial_\mu\psi}{\bar \psi\psi}.
        \end{gather*}
        As in the no derivative case, Lorentz covariance requires that both $A,B$ be proportional to a~linear combination of $I$, $\gamma_5$ and so we may write
        \begin{gather*}
            F=\frac{\left(\partial_\mu\bar \psi\right) ( aI + i b\gamma_5 ) \gamma^\mu \psi}{\bar
            \psi\psi}+\frac{\bar \psi ( cI - i d \gamma_5) \gamma^\mu \partial_\mu\psi}{\bar \psi\psi}   ,
        \end{gather*}
        a result which also satisf\/ies $\mathcal{PCT}$ invariance. Hermiticity of this $F$, and hence current conservation, is satisf\/ied if we have $c=a^*$ and $d=b^*$. Clearly parity invariance is violated if $b \neq 0$; in that case $\mathcal{C}$ invariance requires $b$ to be purely imaginary while $\mathcal{T}$ invariance requires $b$ to be real. The constant $a$ is not constrained by parity but both $\mathcal{C}$ and $\mathcal{T}$ invariance separately require $a$ to be purely imaginary.

Let us consider the special case where each of the discrete
symmetries is individually preserved: $b=0$ and $a=i \epsilon$
with $\epsilon$ a real parameter that controls the strength of the
nonlinearity. Then we may write, using explicitly the on-shell
current conservation condition,
\begin{gather}
                F = 2i\epsilon\frac{\left(\partial_\mu\bar \psi\right)\gamma^\mu\psi}{\bar
                 \psi\psi}   =   -2i\epsilon \frac{\bar \psi \gamma^\mu \partial_\mu\psi}{\bar\psi\psi} .
                  \label{special1}
            \end{gather}

For $\epsilon$ small, one may simplify $F$ in (\ref{special1}) by
solving the nonlinear Dirac equation (\ref{nld}) itera\-tively. To
leading order $(i\gamma^\mu\partial_\mu-m)\psi=0$ which when used
in $F$ gives $F =-2\epsilon m$. Thus to leading order in
$\epsilon$ the nonlinearity (\ref{special1}) is just a mass shift.

 We remark that just as in the no derivative case, we could have imposed $\mathcal{PCT}$ invariance f\/irst and obtained cases which violate Lorentz covariance. However we defer further discussion of Lorentz violating cases to a later section.

\subsection{Two derivatives}\label{section3.3}

There are well-known problems in constructing Lorentz covariant
higher-derivative f\/irst-quan\-ti\-sed theories. Consider a normalised
state,
\begin{gather*}
1 = \int d^3 x \psi^{\dag} \psi   .
\end{gather*}
Applying $\partial \over \partial t$ to both sides gives
\begin{gather*}
 0= \int d^3 x ( \dot{\psi^{\dag}} \psi + \psi^{\dag} \dot{\psi} )  .
\end{gather*}
Now, if the evolution is second-order in time, then one can specify
$\psi(0,x)$ and $\dot{\psi}(0,x)$ independently and that would mean
that the right-hand-side of the above equation need not be zero in general, leading to
a contradiction.

However, in our nonlinear equations, Hermiticity and hence current conservation are ensured by construction and so the above-mentioned problem does not occur. This of course does not guarantee that all other physical quantities will be well-behaved, but it is plausible that that is the case if the higher-order terms are treated perturbatively.

The general structure of the two-derivative nonlinear term, $F \propto I$, without embedded $\gamma$ matrices is
        \begin{gather*}
            F=\frac{a\left(\partial_\mu\partial^\mu\bar \psi\right)\psi+b\bar
            \psi\partial_\mu\partial^\mu\psi+c\left(\partial_\mu\bar \psi\right)\left(\partial^\mu\psi\right)}{\bar \psi\psi}   .
        \end{gather*}
Each numerator/denominator term is separately Poincar\'{e} and
$\mathcal{PCT}$ invariant. However while each term is also
separately parity invariant, $\mathcal{C}$ or $\mathcal{T}$
invariance requires all the coef\/f\/icients~$a$,~$b$,~$c$ to be real.

Current conservation, $F=F^{\dag}$ implies that $b=a^*$ and
$c=c^*$. Thus we conclude that for $a$ not real, both
$\mathcal{C}$ and $\mathcal{T}$ (or $\mathcal{CP}$) are violated.

\section{Plane-wave solutions and dispersion relations}\label{section4}

We wish to construct plane-wave solutions to the nonlinear equations of the previous section. As in the case for the linear theory, we require the solutions to be simultaneous eigenstates of momentum and energy. Let us clarify what this means in the nonlinear theory.

Although we allow the equations to be nonlinear, we keep the fundamental commutation relation between the position and momentum operators, $\left[\hat {\boldsymbol{x}},\hat {\boldsymbol{p}}\right]=i\hbar$. Thus in the Schr\"{o}dinger representation we have $\hat {\boldsymbol{p}}=-i\hbar\boldsymbol{\partial}$ and the momentum eigenvalue is given by
$\hat {\boldsymbol{p}}\psi_p= p \psi_p$.
    Likewise the energy-eigenvalue equation is given by
        $i\hbar\partial_t\psi_E=E\psi_E  $.

    With Lorentz covariance preserved, the method to f\/ind plane-wave solutions is similar to the linear case. We seek solutions of the
    form\footnote{We have set $\hbar=1$.}
    \begin{gather}
        \psi(\boldsymbol{x},t)=e^{-ik.x}u(k) \label{plane}
    \end{gather}
    with $k_{\mu}$ a four vector.

The dispersion relations will be covariantly modif\/ied from that of the linear theory. Consider the nonlinear Dirac equation,
    \begin{gather}\label{9.0a}
        i\partial_t\psi=\left[i\boldsymbol{\alpha}\cdot\boldsymbol{\partial}+\beta m-\beta
        F(\psi)\right]\psi
    \end{gather}
 for the case $F= fI$ where $\alpha^i=\gamma^0\gamma^i$. Substituting the plane wave ansatz into the above equation,
   squaring this and re-arranging gives
    \begin{gather}
                \psi^\dag k^2\psi = \psi^\dag\left[m-f(k_\mu)\right]^2\psi   .\label{9.2}
    \end{gather}
    Thus we have,
    \begin{gather}
        k^2=\left[m-f\left(k^2\right)\right]^2   .  \label{covE}
    \end{gather}
    (Since equation (\ref{9.2}) is covariant, then $f$
    must be also covariant.)

    The solution of (\ref{covE}) requires the explicit form for $f$, the
    nonlinear term. It may also require the explicit form for the
    plane wave solutions which we discuss next. Note that from the above expression, one may view the ef\/fect of the nonlinearity for plane wave states as giving rise to an ef\/fective mass.

Assume $m\neq0$. Then in the rest frame we have from (\ref{plane}), (\ref{9.0a}),
    \begin{gather}
        Mu=\left[\beta m-\beta F(u)\right]u   ,\label{rest}
    \end{gather}
    where the rest energy has been labelled by $M>0$.

    For the case $F \propto I$, the rest frame Hamiltonian is therefore proportional to $\gamma^0 =\beta$ and the eigenstates are as in the linear theory \cite{IZ}.
  These can then be boosted as usual to obtain the general solutions. The net result is similar to the usual spinor solutions of the linear theory but with the ef\/fective mass $M$ in place of the bare mass $m$,
    \begin{gather*}
    E^2 = \boldsymbol{k}^2 + M^2.  
    \end{gather*}
The expression for $M$ in terms of $m$ and the nonlinear parameters can be determined by substituting the rest frame spinors into (\ref{rest}).

    \subsection{Perturbative method}\label{section4.1}
    The procedure of boosting rest frame solutions is valid if Lorentz invariance is a symmetry of the theory.
    If we relax the constraint of Lorentz invariance\footnote{Here we refer to violation of particle Lorentz invariance while keeping observer Lorentz invariance. This can be done by introducing background f\/ields, see Section~\ref{section5}.}, we
    will not be able to use this method to f\/ind the
    energy dispersion relations. Thus we will now introduce a
    method to obtain the energy dispersion relation, to leading order in the nonlinearity, even if we do
    not know the exact plane wave solutions to the theory.

    From (\ref{9.2}), we have
    \begin{gather*}
        E^2=\boldsymbol{k}^2+m^2-2mf+f^2.
    \end{gather*}
    Since the nonlinear term will contain a small nonlinearity
    parameter $\epsilon$, we can explicitly factor it out. That is,
    $f=\epsilon\tilde f$. Then to leading order, we have
    \begin{gather*}
        E^2=\boldsymbol{k}^2+m^2-2\epsilon m\tilde f .
    \end{gather*}
    Now we assume the following,
    \begin{gather*}
        k^\mu = k^{(0)\mu}+O(\epsilon),\qquad
        u(k) = u^{(0)}\big(k^{(0)}\big)+O(\epsilon),
    \end{gather*}
    where $k^{(0)}$ and $u^{(0)}(k^{(0)})$ are the usual
    4-momentum and $u$'s for the linear theory. Thus to leading order
    in $\epsilon$ we have
    \begin{gather}
        E^2 = \left(\boldsymbol{k}^2+m^2\right)-2\epsilon m\tilde
        f\big(u^{(0)}\big(k^{(0)}\big)\big)\!+O\left(\epsilon^2\right)
         = \left(\boldsymbol{k}^2+m^2\right)-2\epsilon m\tilde
        f\big(u^{(0)}(k)\big)\!+O\left(\epsilon^2\right)   .\!\!\! \label{pert-disp}
    \end{gather}
    Note that in the last step, we have replaced $k^{(0)}$ by $k$.
    This is alright because we are dropping terms that are order
    $\epsilon^2$ or higher.

    The perturbative method allows us to f\/ind corrections to the linear theorie's
    energy dispersion relation. We only need to substitute linear plane
    wave solutions into the nonlinear term. Note that the above method works only for the massive theory. If
    we consider the massless limit then we might need to keep terms that are of
    order $\epsilon^2$.

\subsection{Example}\label{section4.2}
We look at an explicit example corresponding to $F \propto I$ and $n=1$ with two derivatives, and obtain the
corresponding expression for the ef\/fective mass $M$ for plane wave
states. Although one can work covariantly with the expressions
(\ref{covE}), it is faster to work in the rest frame, that is by
using (\ref{rest}).

    Consider the nonlinear term when each of the discrete symmetries is preserved,
    \begin{gather*}
        F=\frac{a \partial^\mu \partial_\mu \left( \bar \psi\psi \right)+\left(c-2a\right)\left(\partial^\mu\bar \psi\right)\left(\partial_\mu\psi\right)}{\bar \psi\psi}   .
    \end{gather*}
    Substituting the plane wave solution, the f\/irst term drops out leaving
    \begin{gather*}
        F =  (c-2a) M^2 \equiv \epsilon M^2.
    \end{gather*}
    Thus
    \begin{gather}
        M^2=\left(m-\epsilon M^2\right)^2 .  \label{quartic}
    \end{gather}
    Taking the square root and solving we get
       \begin{gather}
        M=\frac{\mp 1\pm\sqrt{1+4\epsilon m}}{2\epsilon}   . \label{2dersolns}
    \end{gather}
    For the rest energy to be real, we need
    $\epsilon\geq-\frac{1}{4m}$. Let us consider the case where $\epsilon >0$. Then since we have taken $M>0$ by convention, only the following two of the four solutions in (\ref{2dersolns}) are physical:
\begin{gather*}
        M=\frac{\mp 1 + \sqrt{1+4\epsilon m}}{2\epsilon}   .
    \end{gather*}
 In the limit $\epsilon\ll1$, we have
       \begin{gather}
        M = \left\{\begin{array}{l}
                m-\epsilon m^2,\\
                \frac{1}{\epsilon}+m-\epsilon m^2. \label{2solns}
            \end{array}\right.
    \end{gather}
There are therefore two legitimate positive energy solutions for $0<\epsilon \ll 1$. This
is because the equation (\ref{quartic}) is a quartic equation
instead of the usual quadratic which arises when only f\/irst-order
derivatives appear in the Dirac equation. The f\/irst possibility in
(\ref{2solns}) represents a~perturbation to the usual rest mass
and is seen also in the direct perturbative approach of~(\ref{pert-disp}). It results in the dispersion relation
   \begin{gather*}
        E^2 \cong   \boldsymbol{k}^2+m^2-2\epsilon m^3   . 
    \end{gather*}

The other solution in (\ref{2solns}) represents a non-perturbative
mass generation that exists even when $m \to 0$.

\section{Lorentz violating nonlinear equations}\label{section5}

There are various ways of motivating the study of Lorentz violating theories. For example, at short distances space might not be smooth and so dynamical equations might require higher-spatial derivatives to adequately describe the situation. However if one still restricts the time derivatives to f\/irst or second order, to avoid potential causality problems, then clearly one has to give up on Lorentz covariance.

    We will consider nonlinear terms $F$ which simultaneously violate Lorentz invariance \cite{RP2}.
The Lorentz violation will be implemented via constant background f\/ields: in the terminology of \cite{colladay,lehnert} our equations will preserve the observer Lorentz covariance but break the
particle Lorentz symmetry which involves boosting the particles and local f\/ields
but not background f\/ields \cite{colladay,lehnert}.

In this part of the paper we illustrate some of the possibilities
rather than work out all cases as this becomes tedious and is better left for specif\/ic applications.

    As Lorentz violation is constrained by phenomenology to be small \cite{LV,LV1}, we may use per\-tur\-bative methods to determine the corrected dispersion relations.

\subsection{An example: no derivatives}\label{section5.1}

    If the Lorentz violation is described by background {\it vector} f\/ields, then for $F \propto I $ and $n=1$ we may write
     \begin{gather}
        F_1=A_\mu\frac{\bar \psi\gamma^\mu\psi}{\bar
        \psi\psi}+B_\mu\frac{\bar \psi\gamma_5\gamma^\mu\psi}{\bar
        \psi\psi},\label{noder-lv}
    \end{gather}
    where $A_\mu$ and $B_\mu$ are constants; current conservation requires them to be real.

Under a $\mathcal{PCT}$ transformation of the spinors alone
in (\ref{noder-lv}) we have $F \to -F$. Thus we have here our f\/irst
example of $\mathcal{PCT}$ violation associated with Lorentz
violation. However it is possible to maintain $\mathcal{PCT}$
while still violating Lorentz covariance. Consider
    \begin{gather*}
        F_2=A_{\alpha\beta}\frac{\bar \psi\sigma^{\alpha\beta}\psi}{\bar
        \psi\psi}+iB_{\alpha\beta}\frac{\bar \psi\gamma_5\sigma^{\alpha\beta}\psi}{\bar \psi\psi},
    \end{gather*}
    where $A_{\alpha\beta}$ and $B_{\alpha\beta}$ are real background tensor f\/ields. Both current conservation and $\mathcal{PCT}$ invariance are satisf\/ied in this case.

    The dispersion relation for perturbed plane waves can be obtained using the perturbative method, equation~(\ref{pert-disp}). For example, for the case
   \begin{gather*}
        F=A_\mu\frac{\bar \psi\gamma^\mu\psi}{\bar \psi\psi}
    \end{gather*}
    we get $F=\frac{A\cdot k}{m}$. Thus $E^2=\boldsymbol{k}^2+m^2-2A\cdot k$.
  Notice the correction is $O(k)$.

\section{Other cases}\label{section6}

In this section we look at some other examples of nonlinear equations within the class (\ref{form}) such as those with higher nonlinearities, $n \ge 2$, or with $F \propto \gamma_\mu$. We also discuss the Lagrangian approach and some examples of nonlinearities outside the class
(\ref{form}).

\subsection[Lorentz invariant equation with $F\propto I$, $n=2$]{Lorentz invariant equation with $\boldsymbol{F\propto I}$, $\boldsymbol{n=2}$}\label{section6.1}

For simplicity we consider here only cases where there are no derivatives in $F$. An example is given by
\begin{gather*}
    F=\epsilon\frac{\left(\bar \psi\gamma_5\psi\right)^2}{\left(\bar \psi\psi\right)^2}   .
\end{gather*}
It is Poincar\'{e} invariant and invariant under each of the
discrete symmetries while Hermiticity requires $\epsilon$ to be real.
It is easy to verify, using the def\/inition from Section~\ref{section2.5} that $F$ is separable.

\subsection[Lorentz violating equation with $F \propto \gamma_\mu$, $n=1$]{Lorentz violating equation with $\boldsymbol{F \propto \gamma_\mu}$, $\boldsymbol{n=1}$}\label{section6.2}

Here we consider an $F$ that is proportional to
$\gamma^\mu$. Such terms will allow the chiral current to be
conserved, as discussed in Section~\ref{section2.3.2}.
If we exclude derivatives then the simplest possibility is to let the Lorentz index of the gamma matrix contract with that of the background f\/ield $A_\mu$,
\begin{gather*}
    F= i A_\mu\gamma^\mu\frac{\bar \psi\gamma_5\psi}{\bar \psi\psi}.
\end{gather*}
Hermiticity requires the background f\/ield to be real. This $F$
individually breaks all the discrete symmetries and is
$\mathcal{PCT}$ odd! It is separable.

\subsection{Equations from a Lagrangian}\label{section6.3}

There are both advantages and disadvantages in using a
Lagrangian approach. Firstly, a local equation does not
necessarily have a local Lagrangian. Also, even though a Lagrangian might be simple, the resultant equations of motion might look
complicated. On the other hand, it is probably easier to discuss
conservation laws corresponding to symmetries starting from a~Lagrangian. Another possible advantage of a Lagrangian approach
will appear after we look at an example.

Consider the Lagrangian density
\begin{gather*}
    \mathscr{L}=\frac{i}{2}\left[\bar \psi\gamma^\mu\left(\partial_\mu\psi\right)-\left(\partial_\mu\bar \psi\right)\gamma^\mu\psi\right]-m\bar \psi\psi+ \mathscr{L}_{\rm NL}.
\end{gather*}
Suppose, for simplicity, $\mathscr{L}_{\rm NL}$ contains no
derivatives. Then the equation of motion will reduce to
\begin{gather*}
    i\gamma^\mu\partial_\mu\psi-m\psi+\frac{\partial\mathscr{L}_{\rm NL}}{\partial\bar \psi}=0   ,
\end{gather*}
which is a similar to (\ref{nld}) and so we label the last
nonlinear term here as  $F_{\rm E.O.M.}\psi$.
As an example, using,
\begin{gather*}
    \mathscr{L}_{\rm NL}=\frac{\left(\bar \psi A\psi\right)\left(\bar \psi B\psi\right)}{\bar \psi\psi}
\end{gather*}
gives
\begin{gather*}
F_{\rm E.O.M.}\psi =   \frac{\partial\mathscr{L}_{\rm NL}}{\partial\bar
\psi}=\left(\frac{\bar \psi A\psi}{\bar
\psi\psi}\right)B\psi+\left(\frac{\bar \psi B\psi}{\bar
\psi\psi}\right)A\psi-\frac{\left(\bar \psi A\psi\right)\left(\bar
\psi B\psi\right)}{\left(\bar \psi\psi\right)^2}\psi   .
\end{gather*}
Thus we see that a $n=1$ nonlinearity in the Lagrangian will introduce a
mixture of $n=1,2$ terms into the equations of motion.
This then might be one
advantage of the Lagrangian approach: it generates  constrained
complexity from simplicity.

\section{Gauge inequivalence}\label{section7}

It is possible to generate a nonlinear equation from the linear
Dirac equation through a nonlinear gauge transformation~\cite{doebner}. The transformed equation is equivalent to the
original equation in the sense that the probability density is an
invariant. Here we show that the nonlinear terms we have
investigated in this paper cannot be obtained by performing a gauge
transformation on the linear Dirac equation, and so represent
genuine and distinct nonlinear structures.

We def\/ine the following gauge transformation.
\begin{gather*}
    \psi\rightarrow\psi'(x)=e^{i\theta(x)}\psi(x),
\end{gather*}
where $\theta(x)$ is a function of $\bar \psi$'s and $\psi$'s. In
general, we will treat $\theta(x)$ as a $4\times 4$
matrix\footnote{For ease of notation, we will often suppress the
$x$-dependence in $\theta$ and $\psi$.}. We require that
the probability to be invariant under the gauge transformation,
\begin{gather*}
    \psi^\dag\psi\rightarrow{\psi'}^\dag\psi' = \left(e^{i\theta}\psi\right)^\dag e^{i\theta}\psi \equiv \psi^\dag\psi
\end{gather*}
and so $\theta^\dag=\theta$.

Under an inf\/initesimal gauge transformation of the linear Dirac equation we get
\begin{gather}
    \left(1-i\theta\right)\left(i\gamma^\mu\partial_\mu-m\right)\left(1+i\theta\right)\psi  \simeq   0,\nonumber\\
        \left(i\gamma^\mu\partial_\mu-m\right)\psi+\left[\theta,\gamma^\mu\right]\partial_\mu\psi
        -\gamma^\mu\left(\partial_\mu\theta\right)\psi \simeq 0 . \label{g0.3}
\end{gather}
We wish to identify the $\theta$ dependent terms with the nonlinearity $F$ in our nonlinear Dirac equation (\ref{nld})
so we set
\begin{gather*}
    F\psi=\left[\theta,\gamma^\mu\right]\partial_\mu\psi-\gamma^\mu\left(\partial_\mu\theta\right)\psi   .
\end{gather*}
Thus
\begin{gather}\label{g0.5}
    \bar \psi F\psi=\bar \psi\left[\theta,\gamma^\mu\right]\partial_\mu\psi-\bar \psi\gamma^\mu\left(\partial_\mu\theta\right)\psi   .
\end{gather}

We note that equation (\ref{g0.5}) is not symmetric in
$\partial_\mu\psi$ and so this representation of $F$ is
not Hermitian. In order to obtain a symmetric equation, we will repeat
the above steps on the adjoint Dirac equation (this also
removes any ambiguity when taking the adjoint of $\partial_\mu$).

For the adjoint equation, we have for an inf\/initesimal gauge transformation,
\begin{gather}
   0 = \bar \psi\big(i\gamma^\mu\overleftarrow{\partial_\mu}+m\big)+\left(\partial_\mu\bar \psi\right)\left[\gamma^0\theta\gamma^0\gamma^\mu-\gamma^\mu\theta\right]
         +\bar \psi\gamma^0\left(\partial_\mu\theta\right)\gamma^0\gamma^\mu+im\bar \psi\left(\theta-\gamma^0\theta\gamma^0\right)   . \label{g0.7}
\end{gather}
Now the adjoint of (\ref{nld}) is, upon using the Hermiticity
constraint (\ref{herm}),
\begin{gather*}
    \bar \psi\left(i\gamma^\mu\partial_\mu+m\right)-\bar \psi F  = 0   .
\end{gather*}
Thus comparing with (\ref{g0.7}) we label
\begin{gather}\label{g0.8}
    \bar \psi F=-\left(\partial_\mu\bar \psi\right)\left[\gamma^0\theta\gamma^0\gamma^\mu-\gamma^\mu\theta\right]-\bar \psi\gamma^0\left(\partial_\mu\theta\right)\gamma^0\gamma^\mu-im\bar \psi\left(\theta-\gamma^0\theta\gamma^0\right)   .
\end{gather}
Multiplying (\ref{g0.8}) by $\psi$ from the right and adding to (\ref{g0.5}) gives
\begin{gather}
    2\bar \psi F\psi = \bar \psi\left[\theta,\gamma^\mu\right]\partial_\mu\psi-\left(\partial_\mu\bar \psi\right)\left[\gamma^0\theta\gamma^0\gamma^\mu-\gamma^\mu\theta\right]\psi\nonumber\\
\phantom{2\bar \psi F\psi =}{}    -\bar \psi\left[\partial_\mu\left(\gamma^\mu\theta+\gamma^0\theta\gamma^0\gamma^\mu\right)\right]\psi-im\bar \psi\left(\theta-\gamma^0\theta\gamma^0\right)\psi  . \label{g0.10}
\end{gather}
The left hand-side is Hermitian if the constraint (\ref{herm}) on
$F$ is applied. But the adjoint of the right-hand-side is
\begin{gather}
    \left(\partial_\mu\bar \psi\right)\left(\gamma^\mu\gamma^0\theta\gamma^0-\gamma^0\theta\gamma^0\gamma^\mu\right)\psi-\bar \psi\left(\gamma^\mu\theta-\gamma^0\theta\gamma^0\gamma^\mu\right)\partial_\mu\psi\nonumber\\
\qquad{}-\bar \psi\left[\partial_\mu\left(\gamma^0\theta\gamma^0\gamma^\mu+\gamma^\mu\theta\right)\right]\psi+im\bar \psi\left(\gamma^0\theta\gamma^0-\theta\right)\psi   . \label{g0.11}
\end{gather}
Comparing (\ref{g0.10}) and (\ref{g0.11}), we require
\begin{gather*}
    \gamma^0\theta\gamma^0=\theta   \equiv     \left[\theta,\gamma^0\right]=0   .
\end{gather*}
Then (\ref{g0.10}) becomes
\begin{gather}\label{g0.13}
    2\bar \psi F\psi=\bar \psi\left[\theta,\gamma^\mu\right]\partial_\mu\psi-\left(\partial_\mu\bar \psi\right)\left[\theta,\gamma^\mu\right]\psi-\bar \psi\left[\partial_\mu\left\{\theta,\gamma^\mu\right\}\right]\psi   .
\end{gather}

So far we have deduced two constraints on $\theta$,
\begin{gather*}
    \theta = \theta^\dag   , \qquad
    \left[\theta,\gamma^0\right] = 0   ,
\end{gather*}
coming respectively from the invariance of the probability
density and Hermiticity. These are {\it necessary} constraints for
a nonlinear equation generated by gauge transformation to be
equiva\-lent to a theory of our general class, but one must still
check if any candidate solution,~$\theta$, is actually a solution,
that is, suf\/f\/iciency is not guaranteed by~(\ref{g0.13}).

\subsection{Lorentz invariant case}\label{section7.1}

We will now look at the constraint from Poincar\'{e} invariance.
Recall that we need $S^{-1}F'S\equiv F$ under
$\psi\rightarrow\psi'=S\psi$. The l.h.s.\ of (\ref{g0.13}) is clearly
invariant while the r.h.s.\ transforms into
\begin{gather}
    \bar \psi S^{-1}\left[\theta',\gamma^\mu\right]\Lambda^\nu_\mu S\partial_\nu\psi-\left(\partial_\nu\bar \psi\right)S^{-1}\Lambda^\nu_\mu\left[\theta',\gamma^\mu\right]S\psi
 -\bar \psi\left[\partial_\nu S^{-1}\Lambda^\nu_\mu\left\{\theta',\gamma^\mu\right\}S\right]\psi   . \label{g0.14}
\end{gather}
Comparing (\ref{g0.14}) with (\ref{g0.13}), we get
\begin{gather*}
                    \left[S^{-1}\theta'S,\gamma^\nu\right] \equiv\left[\theta,\gamma^\nu\right],\qquad
                    \left\{S^{-1}\theta'S,\gamma^\nu\right\} \equiv\left\{\theta,\gamma^\nu\right\}.
\end{gather*}
Thus we have the constraint
\begin{gather*}
    S^{-1}\theta'S=\theta,
\end{gather*}
which for an inf\/initesimal Lorentz transformation gives
\begin{gather*}
    \theta'-\frac{i}{4}\omega^{ab}\left[\theta',\sigma_{ab}\right]=\theta   .
\end{gather*}
Therefore in total we have 3 constraints,
\begin{gather*}
            \mbox{constraint 1:} \quad \theta^\dag=\theta,\\
            \mbox{constraint 2:} \quad \big[\theta,\gamma^0\big]=0,\\
            \mbox{constraint 3:} \quad \theta'-\frac{i}{4}\omega^{ab}\left[\theta',\sigma_{ab}\right]=\theta   .
\end{gather*}
From constraint~2, $\theta$ must be proportional to $I$ or
$\gamma^0$. If $\theta\propto I$, then all constraints are satisf\/ied but for $\theta\propto\gamma^0$, we cannot satisfy
constraint~3: Let $\theta=g\gamma^0$, where $g$ is a scalar
function of the wavefunctions. Then the Poincar\'{e} transformed
$\theta$ is given by $\theta'=g'\gamma^0$. Substituting this into
the left-hand-side of constraint~3, we get
\begin{gather*}
    g'\gamma^0-\frac{i}{4}g'\omega^{ab}\left[\gamma^0,\sigma_{ab}\right]=g'\gamma^0-\frac{i}{4}g'\omega^{ab}\left(\gamma_a-\gamma_b\right)  .
\end{gather*}
Since $\omega^{ab}\left(\gamma_a-\gamma_b\right)$ is
non-zero, the result is not proportional to $\gamma^0$ and so $\theta \propto \gamma^0$ does not satisfy constraint 3. Thus we conclude that $\theta$ can only be proportional to $I$.

Hence with $\theta \propto I$, equation (\ref{g0.13}) becomes
\begin{gather}\label{g0.18}
    \bar \psi F\psi=-\frac{1}{2}\bar \psi\left[\partial_\mu\left\{\theta,\gamma^\mu\right\}\right]\psi=-\bar \psi\left(\partial_\mu\theta\right)\gamma^\mu\psi=-j^\mu\partial_\mu\theta    .
\end{gather}

Consider the specif\/ic case where $F$ is proportional to $I$. Writing $F=fI$, we deduce from~(\ref{g0.18}) that
\begin{gather}
           f = -\frac{\left(\partial_\mu\theta\right)j^\mu}{\bar \psi\psi}   . \label{g0.21}
\end{gather}
Remember that $\theta$ is a function of $\bar \psi$'s and $\psi$'s,
and recall our condition~(\ref{scale}):  we see therefore that $\theta$
must be invariant under a scaling of the wavefunction. As long as
the nonlinearities cannot be expressed in the form shown in~(\ref{g0.21}), we can be sure that they cannot be obtained by
performing a gauge transformation on the linear Dirac equation. In
particular we conclude that the Lorentz covariant nonlinear Dirac
equations we have explicitly studied in this paper are
not gauge equivalent to the linear Dirac equation.

Now consider the class of nonlinearities where $F$ is
proportional to $\gamma^\mu$. We let $F=f_\mu\gamma^\mu$, where
$f_\mu$ are functions of $\bar \psi$'s and $\psi$'s. Then
(\ref{g0.18}) becomes
\begin{gather}\label{g0.25}
    f_\mu\bar \psi\gamma^\mu\psi=f_\mu j^\mu=-j^\mu\partial_\mu\theta.
\end{gather}
Therefore if $f_\mu$ cannot be expressed as a total
derivative of a scale-invariant $\theta$ function like~(\ref{g0.25}) then those nonlinear
structures proportional to $\gamma^\mu$ cannot be obtained from the
linear Dirac equation by a gauge transformation. In particular the
cases we considered in Section~\ref{section6.2} are safe.

\subsection{Lorentz violating cases}\label{section7.2}

Finally let us consider the case where $F$ is Lorentz violating.
We have constructed our Lorentz violating terms by introducing a
constant background f\/ield $A_\mu$ (independent of the
wavefunction). We may write $F$ as $A_\mu G^\mu$ where $G^\mu$ is
the nonlinear factor which may  be proportional to $I$,
$\gamma^\mu$ etc.

Could the Lorentz violating examples we have considered be obtained by a nonlinear gauge transformation of the linear Dirac equation with or without Lorentz violation?
The linear Dirac equation to start with would now be of the form
 \begin{gather*}
    \left(i\gamma^\mu\partial_\mu-m\right)\psi+ LV \psi=0,
\end{gather*}
where $LV$ is a state-independent Lorentz violating term, if it is
not zero (we assume that $LV$ does not have free derivatives that
act to the right on $\psi$). Gauge transforming this equation with
a state-dependent but Hermitian $\theta \propto I$ can generate at
most Lorentz covariant nonlinearities. So consider the other
possibility, $\theta \propto \gamma_0$. Then one would generate
Lorentz violating nonlinearties and on the right-hand-side of
(\ref{g0.3}) there would be an additional term $\sim [LV,
\gamma_0]$. Now if we write $\theta=\bar \theta\gamma^0$,
(\ref{g0.13}) becomes
\begin{gather}\label{g0.26}
    \bar \psi F\psi=\bar \theta\psi^\dag\gamma^i\partial_i\psi-\bar \theta\big(\partial_i\psi^\dag\big)\gamma^i\psi-\left(\frac{\partial}{\partial t}\bar \theta\right)\bar \psi\psi   .
\end{gather}
The f\/irst observation is that in order to write the
right-hand-side in covariant form we need to introduce background
tensor (for the f\/irst two terms) and vector (for the last term)
f\/ields. Also from the structural form of our $F$ (\ref{form}), we
see by comparing both sides of (\ref{g0.26}) that $\theta$ must be
invariant under scaling of the wavefunction. The examples we have
explicitly discussed in this paper therefore do not fall under the
category of nonlinearities described by (\ref{g0.26}). For
example, with $F=fI$, (\ref{g0.26}) becomes
\begin{gather}
    f=\bar \theta\left[\frac{\psi^\dag\gamma^i\partial_i\psi-\left(\partial_i\psi^\dag\right)\gamma^i\psi}{\bar \psi\psi}\right]-\dot{\bar \theta}   ,
\end{gather}
which means having at least $n=2$ and a simultaneous use of tensor and vector f\/ields: these are necessary conditions for the nonlinearity to be obtained through a Lorentz violating gauge transformation of the usual linear Dirac equation.

\section{Discussion}\label{section8}

In \cite{RP2} it was suggested that fundamental quantum nonlinearities might be related to potential Lorentz violation \cite{colladay,lehnert,LV,LV1}. This current paper is a step towards a quantitative study of the suggestions in \cite{RP2}. We have discussed a framework for systematically constructing nonlinear Dirac equations, at the quantum mechanical level, that satisfy other conventional properties such as Hermiticity,  Poincar\'{e} invariance and $\psi \to \lambda \psi$ invariance although, as shown, even those can be relaxed.

We gave several examples of such equations, dif\/ferent in structure from those studied pre\-viously in the literature, and discussed their properties. We also demonstrated that our equations were not gauge equivalent to the linear Dirac equation. More explicit examples of our class of nonlinear Dirac equations
may be found in~\cite{NP1} and their non-relativistic limit is studied in~\cite{NP3}.

As mentioned in Section~\ref{section1}, one application of such equations is to study neutrino oscillations~\cite{NP2} which would be an ideal probe of quantum nonlinearities, with of without a simultaneous Lorentz violation~\cite{RP2}. Other examples we hope to study with the nonlinear equations are $\cal{CP}$ violation and dark matter/energy. In this regard, it would be useful to obtain non-plane-wave solutions to our nonlinear equations, similar to what has been done for simpler polynomial-type nonlinear Dirac equations in~\cite{NLDS,NLDS1}.

A number of authors had argued that nonlinear quantum evolution of states within the standard kinematical framework of quantum theory would lead to pathologies. However, on closer examination, such attempts at ``no go" theorems were seen to require one or more assumptions that are not very obvious on physical grounds; for detailed critiques and citations to the literature the interested reader is referred to \cite{Holman,Holman1}.

We have kept open the possibility that the nonlinearities we proposed might be fundamental, ef\/fective or only phenomenological.
Of course there is less contention if the nonlinearities are only an approximate representation of more complex underlying dynamics; in any case, from a~Wilsonian perspective, one deals in physics with a sequence of  approximate theories.

Ef\/fective or phenomenological nonlinear equations are quite common in the non-relativistic domain \cite{book1,book2} and there are also a few examples of phenomenological relativistic nonlinear equations \cite{Misc,Misc1,Misc2}. As another possibility of the latter case, we note that some condensed matter systems have (linear) relativistic-looking equations for their quasi-particles \cite{graphene}: these are surely approximations to nonlinear equations.

\pdfbookmark[1]{References}{ref}
\LastPageEnding

\end{document}